# A Colossal Electroresistance response, accompanied by metal-insulator transition, in a mixed-valent vanadate

Rafikul Ali Saha[1], Abhisek Bandyopadhyay[1], Irene Schiesaro[2], Carlo Meneghini[2], and Sugata Ray[1]

[1]*School of Materials Sciences, Indian Association for the Cultivation of Science, 2A & 2B Raja S. C. Mullick Road, Jadavpur, Kolkata 700032, India and*
[2]*Dipartimento di Scienze, Universitá Roma Tre, Via della Vasca Navale, 84 I-00146 Roma, Italy*

## Abstract

Colossal electroresistance (CER) in manganites, *i.e.*, a large change in electrical resistance under the influence of either an applied electric field or an applied electric current, has often been described as complimentary to the colossal magnetoresistance (CMR) effect. Mixed valent vanadates with active $t_{2g}$ and empty $e_g$ orbitals, unlike manganites, have not naturally been discussed in this context, as double exchange based CMR is not realizable in them. However, presence of coupled spin and orbital degrees of freedom, metal-insulator transition (MIT) accompanied by orbital order-disorder transition, etc., anyway make the vanadates an exciting group of materials. Here we probe a Fe-doped hollandite lead vanadate $PbFe_{1.75}V_{4.25}O_{11}$ (PFVO), which exhibits a clear MIT as a function of temperature. Most importantly, a giant fall in the resistivity, indicative of a CER, as well as a systematic shift in the MIT towards higher temperature are observed as a function of applied electric current. Detailed structural, magnetic, thermodynamic and transport studies point towards a complex interplay between orbital order/disorder effect, MIT and double exchange in this system.

## Introduction

The orbital degrees of freedom (ODF) of the correlated 3*d* transition metal oxides (TMOs), with metal-oxygen octahedral building blocks, has been one of the most fascinating condensed matter research topics for decades now. It offers a plethora of novel and extraordinary physical properties, e.g. metal-insulator transition (MIT), colossal magneto and electroresistance, magnetization reversal, electronic phase separation, ferroelectricity, spin-glass freezing, spin, charge, orbital ordering, etc[1-4], where the orbital quantum number determines the electron density distribution and hence, provides a link between magnetism and structure of the chemical bonds[5]. The ODF in a solid can be manipulated via (i) the orbital-lattice Jahn-Teller (JT) coupling, (ii) the magnetic exchange interactions between the orbitals, (iii) the spin-orbit coupling, and (iv) extrinsic effects due to ionic size mismatch of cations, external applied pressure, presence of lone pair, etc[5-7]. In case of strong JT interaction (active orbitals of $e_g$ symmetry) the completely uncorrelated spin and orbital degrees of freedom could facilitate very different magnetic and structural transition temperatures within larger lattice distortions[7]. On the other hand, the $t_{2g}$-active orbitals (Ti,

V, Ru, etc.,) possess weak JT distortions, and consequently, strong competition between the spin interactions and the orbital ordering/disordering phenomena may produce nearly coinciding magnetic and structural transition temperatures[7]. Therefore, the lattice driven classical orbital picture, as in manganites, is contrasted by the quantum behaviour of orbitals in frustrated exchange interaction models of the early 3$d$ TMOs.

Naturally, vanadates are considered to be model canonical systems for investigating a rich variety of outstanding magnetic and electrical behaviors originating from the interconnected knot of spin, charge, orbital and lattice correlations of the V $t_{2g}$ electrons[8-13]. The $t_{2g}$ valence electrons of the V ions (V$^{3+}$ : $t^2_{2g}$/V$^{4+}$: $t^1_{2g}$) undergo significantly lower noncubic crystal field splitting, as compared to that of the $e_g$ manifold in isostructural manganites, resulting in the comparable energy separation of orbital levels to the spin interactions[7,11]. Hence, the interplay between orbital and spins, (not only the intersite exchange interaction but also the intra-atomic spin-orbit interaction) becomes more pronounced within orbitally active V $t_{2g}$ electrons[14-17] of the vanadates, causing multiple magnetic and orbital ordering transitions, accompanying with the orbital ordering/disordering-induced MIT[11-13].

In such a backdrop, another extremely fascinating series of mixed-valence R-block-type vanadates of general chemical formula $AV_6O_{11}$ (A = K, Na, Sr, Pb)[18-21] appear truly promising for exploring a variety of intriguing magnetic and electronic properties. The monovalent A-site ion (K$^+$, Na$^+$), as well as the divalent ones (Sr$^{2+}$, Pb$^{2+}$), all exhibit a common structural phase transition from high temperature centrosymmetric ($P6_3/mmc$) to low temperature non-centrosymmetric ($P6_3mc$) structure, triggered by cationic size mismatch and lone pair effect, respectively[18-21]. The magnetic ordering temperatures ($T_C$ = 64.2, 66.8, 73, and 90 K for the A = Na, K, Sr, and Pb respectively) are quite apart from the structural transition temperatures but are intimately accompanied with the slope change in the respective electrical resistivity versus temperature curves[21-24], which might be indicative of significant spin-lattice-orbital coupling of the orbitally active V $t_{2g}$ electrons[7] at the onset of magnetic transition. Surprisingly the temperature dependent electrical resistivity measurements revealed a semiconducting/insulating behavior for the $A^{2+}V_6O_{11}$ (A = Sr$^{2+}$ and Pb$^{2+}$) in contrast with the metallic transport of the monovalent Na, K members[21,24]. Clearly the direct difference between the divalent and monovalent A cation members are, (1) difference in effective V$^{3+}$/V$^{4+}$ ratio (V$^{3+}$:V$^{4+}$ = 2:1 for $A^{2+}$ and 1:1 for $A^+$)[21,24], and (2) the difference in the A cation size and the resultant structural distortion. One option to independently probe these parameters could be to dope the V atoms in a way so that the V$^{3+}$/V$^{4+}$ ratio are manipulated, keeping the A cation unchanged. Therefore, it becomes an exciting proposition to replace some proportion of V$^{3+}$ of a $A^{2+}V_6O_{11}$ compound by the isovalent Fe$^{3+}$ species, and taking the average V oxidation state close to ~3.5, i.e., similar to $A^+$ members, thereby making the doped compounds a fertile ground for investigating new physics. However, such a strategy has its own weaknesses because the dopant ion will have different electronic structure (e.g. active $eg$ orbital for Fe$^{3+}$, unlike V$^{3+}$) and consequent differences in spin, orbital, and lattice degrees of freedom. Moreover, such changes around the dopant would also alter the magnetic exchange pathways and as a result, the global magnetic order may get interrupted by the local intervention of the dopant ions.

PbFe$_{1.75}$V$_{4.25}$O$_{11}$ (PFVO) becomes important exactly in this context, where PFVO can be treated as an effectively hole-doped counterpart of the parent PbV$_6$O$_{11}$, with the resultant V$^{3+}$:V$^{4+}$ ratio being nearly 1:1. Even though the parent PbV$_6$O$_{11}$ is an insulator, the first exciting observation that we record in PFVO is a stunning development of a resistivity saturation at around the magnetic transition temperature (~85 K), followed by an entry to a metallic state with further lowering of temperature. This indicates that metallic transport in the system is predominantly determined by the effective V$^{3+}$:V$^{4+}$ ratio, because Fe-doping ensures only this change. It is interesting to note that the overall magnetic behavior as well as the spin glass freezing temperature only marginally change with doping, which confirms that replacement of V$^{3+}$ by Fe$^{3+}$ in the structure does not significantly affect the exchange pathways. However, the most astonishing observation is the existence of a colossal electroresistance in this compound, exactly around the onset of MIT, as a function of applied electric current, which has never been anticipated for a $t_{2g}$-active lower transition metal system. Absence of any magnetoresistance in the system, even around the MIT, clearly establishes that the observed CER is not connected with any magnetic field driven transition of the ground state. The relative competition between the Jahn-Teller (JT) distortion mediated orbital ordering favoring electron localization and the V$^{3+}$-O-V$^{4+}$ double exchange driven orbital disordering facilitating movements of the doped holes, turns out to be at the root of all the emerged unusual behaviours.

## RESULTS AND DISCUSSIONS

***Structural analysis.*** Synchrotron x-ray diffraction data (Fig. 1(a)), collected from polycrystalline powder at both room temperature and 15 K, ensures pure single phase of the sample without any structural phase transition down to the lowest measured temperature, and consequently, satisfactory structural refinement has been performed by considering a non-centrosymmetric hexagonal *P*6$_3$*mc* space group, which is consistent with the previous literature[25,26]. The refined structural parameters including lattice constants, atomic positions, site occupancy, along with the goodness factors are listed in Table-S1. The refined crystal structure is shown in Fig. 1(b) which shows that there are four distinct crystallographic sites (*M*(1), *M*(2), *M*(3) and *M*(4)) for the transition metal (TM) cations, among which *M*(1) forms Kagome network in the *a-b* plane by edge-sharing *M*(1)O6 octahedra. Our structural analysis clearly indicates that Fe completely substitutes V at the trigonal bipyramidal *M*(4)-site, both *M*(2)- and *M*(3)-sites of the *M*(2)*M*(3)O$_9$ dimer are randomly occupied by the V/Fe, while the *M*(1)-site is solely occupied by V (see Table-S1). It is well known that the chemical order at the local scale may differ from the long range order probed by crystallography[27-29], therefore the Fe *K*-edge EXAFS measurement has been carried out on this sample in order to probe the local atomic structure and chemical order around Fe atoms. The Fe *K*-edge best fit and the fitted structural parameters are presented in Fig. S2 and Table-S2. The EXAFS analysis confirms that the average local structure around Fe ions is consistent with the crystallographic structure derived from the Rietveld refinement of XRD data (See details in SM).

Next it is observed that due to stereochemically active Pb$^{2+}$ 6$s^2$ lone pair, large structural distortions in terms of tilting, rotation and elongated/compressed TM-oxygen bond distances of the TM-oxygen polyhedra (see Figs. S1(a)-(d)), as well as large structural asymmetry via significant off-centering of the metal cations (see Figs. 1(c)-(d), and Table-I) are present in the structure. As displayed in Fig. 1(c) and Table-I, it is evident that the V(1)O$_6$ octahedra possess $Z_{out}$ type Jahn-Teller (J-T) distortion through local Z-axis elongation of the apical V(1)-O bond lengths. On top of it, the 300 K refined crystal structure illustrates that the V(1) atom also gets significantly off-centered (see Fig. 1(c) and Table-I) through the creation of dissimilar (i.e. long and short) V(1)-O distances in each of the three V(1)-O pairs of the V(1)O$_6$ octahedra. Most interestingly, such off-centering of V(1) cation completely disappears in the basal plane at 15 K (see Table-I), while the $Z_{out}$ J-T distortion persists down to 15 K (Table-I). On the other hand, the M(2)-O and M(3)-O distances clearly indicate that both the M(2) and M(3) octahedra remain J-T inactive, but, significant shifting of the M(2) and M(3) cation centers is envisioned in the direction opposite to that of Pb-center (see Fig. 1(d) and Table-I). Furthermore, the V(1) ions of the edge-shared V(1)O$_6$ octahedra, forming regular kagome lattice in the *a-b* plane, get displaced from their respective octahedral centers as a result of large octahedral distortions and structural asymmetry at the *P63mc* space group, thereby, forming kagome trimers consisting of two distinct sets of corner-shared V(1) equilateral triangular networks, one having longer V(1)-V(1) and the other with shorter V(1)-V(1) distances (Fig. 1(e)).

***Estimation of transition metal's valance state.*** Next we performed the Fe and V K-edge x-ray absorption near-edge structure (XANES) spectroscopy measurements to authenticate the respective cation's oxidation states, as well as carried out the bond valance sum (BVS) calculations to affirm the site-specific valance states of V in the present study. A detailed discussion on both the XANES and BVS analysis have been summarised in the SM. Interestingly Fe substitution brings into significant changes in the V-valance distributions in their respective sites (see Table-II), in contrast to the other members of the $AV_6O_{11}$ series.

***Magnetization and heat capacity.*** Now we focus on the magnetic properties of PFVO. The temperature variations of the *dc* magnetic susceptibility χ have been measured in the zero field-cooled (ZFC), field-cool-cooling (FCC) and field-cool-heating (FCH) protocols under three different applied magnetic fields (100 Oe, 500 Oe and 2000 Oe), and the 100 Oe χ(T) data are presented in Fig. 2(a). A steep upturn is evident near around 85 K in both the observed ZFC and FC data (see $d\chi/dT$ versus T, shown in the inset to Fig. 2(a)). In order to comment on the nature of this transition the temperature dependent heat capacity ($C_p$ versus T) has been measured (see Fig. S4(a)), where the absence of any anomaly in the $C_p$ data strongly refutes long-range coherent magnetic order[31-34]. However, the temperature dependent inverse susceptibility variation (1/χ versus T) under 100 Oe field shows a sharp downturn below about 275 K, and a gradual smearing out of the same with increasing applied fields (see Fig. 2(b)), signifying the presence of short range ferromagnetic correlations within the paramagnetic background. Isothermal remnant magnetization (IRM) at 15 K and the temperature dependent ac-susceptibility at a set of discrete frequencies ranging from 21 Hz to 1111 Hz, have been carried out using conventional method. It is evident that the $M_{IRM}$ undergoes a slow decay (Fig. S4(b)), and the obtained $M_{IRM}$ versus t data is further fitted using the logarithmic function, $M_{IRM}(t)/M(0) = A_0 - S*\ln(1 + t/t_0)$, as illustrated in Fig. S4(b), which is a signature of magnetic materials having hysteretic magnetization and/or glassy dynamics[34-36]. Further, the temperature variations of both the in-phase and out-of-phase susceptibilities χ′ (ω, T) and χ′′(ω, T) (see Fig. 2(c)) show a well-defined maximum at~ 85 K, which has a clear frequency dispersion, thereby affirming the spin glass ground state[37-39]. Considering the two distinct corner-shared equilateral V(1) triangular units within the kagome network (see Fig. S4(c)), competing FM and AFM spin arrangements as well as disordered distributions of V and Fe within $M(2)$ and $M(3)$ sites, the spin glass ground state of this compound below 85 K can be explained. Apart from the short range FM correlation and the spin-glass magnetic ground state, a distinct thermal hysteresis between the FCC and FCH magnetization data is observed (see Fig. 2(d)). Interestingly, the FCC and FCH data do show clear thermal hysteresis at least at two temperature regions (∼ 95 - 62 K and ∼ 45 - 30 K, as highlighted in Fig. 2(d)). Further, the thermal hysteresis around $T_C$ is relatively widely extended in the temperature span, and the FCC and FCH lines cross each other at two temperatures ∼ 60 K and 48 K. Such divided hysteresis loop could be an indication of multiple first order phase transitions[40].

*Electrical Resistivity.* Next we investigate the temperature dependent electrical resistivity ($\rho$-$T$) of this sample at various applied currents, and the measured $\rho$-$T$ data have been presented in Fig. 3(a) for both the cooling and heating cycles. The higher temperature insulating nature ($d\rho/dT < 0$) of transport is followed by the observation of resistivity saturation in the form of flattened $\rho(T)$ curve at the onset of magnetic transition for the lowest current (1 $\mu$A) measurement, and finally a sharp metal-insulator transition (MIT) by a strong suppression of resistivity. In PFVO, the mixed $V^{3+}/V^{4+}$ valancy and the completely random distribution of V and Fe in the $M(2)$- and $M(3)$-sites of the crystal structure introduce various magnetic exchange interaction pathways through face-, edge- and corner-shared polyhedral connectivity in between the $M(i)$ ($i$ = 1 to 4) sites. This facilitates the microscopically intrinsic electronic phase separation within the system. Like perovskite manganites, such electronic inhomogeneity in PFVO could be associated with the competition between the correlated-metal and the Mott-insulating phases[41]. With decreasing temperature from 300 K, a continuous increase in the resistivity till the appearance of at $\rho(T)$ curve signifies insulating transport in that specific temperature range. Due to tilting distortions of the $V(1)O_6$ octahedra and the consequent local $Z$-axis elongation (see in Fig. 4(b)), a collective $Z_{out}$ Jahn-Teller (J-T) distortion gets stabilized at the V(1) center at higher temperature. This J-T effect eventually lifts degeneracy of the $t_{2g}$ orbitals of the $V^{4+}$ ($t^1_{2g}$) cation in the V(1)-site (Fig. 4(b)), thereby resulting in the half filled low-lying doubly degenerate $d_{xz}$, $d_{yz}$ orbitals and higher energy empty $d_{xy}$ level. As a result of which, the $t^1_{2g}$ electron occupies in any of the two degenerate levels ($d_{zx}/d_{yz}$) maintaining an orbital disorder situation. But interestingly, on top of the J-T distortion, the V(1) cation shows a significant off-centering movement in the $xy$ basal plane, creating further splitting in the doubly degenerate $d_{zx}/d_{yz}$ orbitals and giving rise to the orbital ordering of the $t^1_{2g}$ electron. So, the $t^1_{2g}$ electron occupies the lower energy $d_{zx}(d_{yz})$ orbital keeping the higher energy $d_{yz}(d_{zx})$ level empty providing an energy gap in between them, and hence an orbital ordered insulating (OOI) state is emerged at the high temperatures. Besides, the observed insulating resistivity within the given temperature region could be interpreted in terms of the enhanced carrier scattering by localized spin fluctuations in the OO state and resultant spatial randomness of the transfer interaction ($t_{ij} \sim \cos(\Delta\theta_{ij}/2)$), with $t_{ij}$ being transfer integral between the neighboring atomic sites and $\Delta\theta_{ij}$ is the relative angle of the local spins) i.e., off- diagonal disorder effect[42]. Following insulating resistivity the sample enters a bad metal state through orbital disordering and shows a resistivity saturation (flattening of $\rho$-$T$ curve) within the Ioffe-Regel criterion, where the resistivity does not increase beyond a certain saturation value corresponding to the comparable electron mean free path and inter atomic distances[43-45]. It should be noted that there is competition between the J-T distortion plus off-centering assisted orbital ordering (OO) of $V^{4+}$ ion (Fig. 4(b)-(c) and Fig. 1(c) along with Table-I) in the V(1) kagome layer and the $V^{3+}$-O-$V^{4+}$-O-$V^{3+}$ double exchange (DE) pathway driven ferromagnetic interactions within V(1) and V(2)/V(3) sites (see Fig. 4(d)). Therefore, one should point towards the fact that at the onset of ferromagnetic transition ($T_c$) the DE stabilization overcomes the energy gap due to orbital ordering phenomenon, thereby facilitating electron conduction and hence MIT through orbital disordering[11-14,46,47]. This MIT is further accelerated with temperature lowering due to the disappearance of V(1)-off-centering movement in the basal plane (see Fig. 4(c) and Table-I) and the consequent suppression of energy level splitting between the $d_{zx}$ and $d_{yz}$ orbitals. The $t^1_{2g}$ electron therefore prefers to be arranged in degenerate $d_{zx}/d_{yz}$ orbitals at low temperatures, favoring orbital disordering and consequent metallic band structure.

***Current induced change in electrical resistivity.*** Now we turn our focus on the temperature dependent electrical resistivity measurements under different applied electrical currents (1 μA, 10 μA, 20 μA, 30 μA and 70 μA). The measured $\rho$-$T$ data are illustrated in Fig. 3(a). In addition to the MIT, a few intriguing aspects regarding the observed current-induced spectacular changes in electrical resistivity are summarized as follows:
1) The occurrence of a giant square shaped hysteresis loop between the heating and cooling cycles of the $\rho$-$T$ curves measured above a critical value of the applied current (>20 μA).
2) With increasing applied current (at and above 20 μA), two more extra features appear at around 52 K and 40 K in the $\rho$ ($T$) curve.
3) The gradual decrease in the maximum value of resistivity at the onset of MIT, and simultaneously a systematic shift in the MIT temperature towards higher values with increasing applied electric current (shown in Figs. 3(a)-(c)).

It is already discussed that at the low current region, the relative competition between the orbital ordering and the ferromagnetic DE interaction is responsible for the observed MIT. So, even at a small applied *dc* current (field) there is melting of OO state in the vicinity of $T_c$, giving rise to the insulator-metal transition through a huge reduction in the measured resistance[48-50]. Now with increasing applied current the electric field above a critical value further triggers the DE interaction, and therefore, helping to destabilize the orbital ordering in this sample. Hence, the appearance of `resistivity saturation' and the consequent MIT start to shift systematically towards higher temperatures (Figs. 3(a) and (b)) with higher values of applied current. Such a peculiar trend is the outcome of a delicate balance between double exchange, electric field, thermal energy, and orbital ordering/disordering. The observed systematic decrease in the measured maximum resistivity at the onset of MIT by orders of magnitude as well as the gradual shift of the metal-insulator transition towards higher temperatures with increasing applied current can be aptly called the colossal electroresistance (CER) phenomenon. The CER-related observations on the application of electric current (field) could emerge collectively from (1) the decrease in spin-disorder scattering, (2) the growth of the domains giving rise to better metallic path, as explained by double exchange mechanism where metallic conduction is coupled with FM interactions, (3) increasing mean-free path of electrons, and (4) growth of aligned domains in applied field giving rise to decrement in the probability for domain-wall scattering[51]. Therefore, it is quite natural to propose that the applied external electric field stimulates melting of V(1) orbital ordering by enhancing $V^{3+}$-O-$V^{4+}$ DE interaction strength between the V(1) and V(2)/V(3)- sites (Fig. 4(d)), thereby causing delocalization of intrasite holes through the generation of mobile electronic charges, and ensuring the MIT and CER phenomena in PFVO.

Finally, the observed strong hysteresis between the heating and cooling cycles of the resistivity curves above a critical current (> 20 $\mu$A, Figs. 3(a) and (c)) likely originates due to the coexistence of two competing phases, namely, orbital ordered Mott insulator and correlated FM metal, thereby identifying the MIT as first order in nature[52-54]. The presence of large hysteresis or such metastable states in the course of melting transition of the orbital-ordered state[52] would also favor such a resistivity switching. Furthermore, with increasing applied current (at and above 20 $\mu$A), two more extra features appear at around 52 K and 40 K in the $\rho(T)$ curves (Fig. 3(a)), supporting the existence of multiple first order phase transitions, which is in agreement with the observations from temperature dependent *dc* susceptibility measurements (Fig. 2(d)).

But surprisingly, none of the applied magnetic fields (up to 8 Tesla) does show any change in the electrical resistivity, negating any possibility of the colossal magnetoresistance to be present in this sample (see Fig. S5). So we should affirm that the emergence of colossal electroresistance property in this PFVO compound has no electronic connection with CMR.

## CONCLUSION

In this work the structural, electronic, magnetic and transport properties of polycrystalline PbFe$_{1.75}$V$_{4.25}$O$_{11}$ sample are experimentally studied in detail. The relative proportion of the mixed vanadium valancy (V$^{3+}$ with respect to V$^{4+}$ upon ~ 30% Fe-doping), revealed from the V *K*-edge XANES measurement, places this particular composition of the *A*V$_6$O$_{11}$ series as an electronic analogue to the hole-doped manganites. Our detailed electrical transport measurements clearly affirm the metal-insulator transition (MIT) and colossal electroresistance (CER) phenomena in the system. The double exchange driven ferromagnetic interactions and the externally applied electric field together help the delocalization of the charge carriers, which eventually act against the structural distortion driven stabilization of orbital ordering, thereby promoting a subtle competition between the orbital ordered and disordered states, which is at the root of all the observed spectacular electrical properties in this compound. On the other hand, our in-depth *dc* and *ac* magnetization measurements confirm the spin-glass magnetic ground state.

## ACKNOWLEDGEMENT

R.A.S and A.B thank CSIR, India, and IACS for fellowships. S.R thanks DST for funding (CRG/2019/003522). The authors also thank Indo-Italian POC for support to carry out experiments in Elettra, Italy, and Laboratory for Materials and Structures, Tokyo, Japan for providing experimental facilities through collaborative research project. S.R also thanks the Technical Research Center (TRC) of IACS for experimental facilities.

TABLE I. V(1)-/$M$(2)-/$M$(3)-/Fe(4)-O bond distances at 300 K and 15 K

| Metal-O | Bond length (Å) 300 K | Shift 300 K | Bond length (Å) 15 K | Shift (Å) 15 K |
|---|---|---|---|---|
| V(1)-O(4) | 2.09 | 0.14 | 2.10 | 0.17 |
| V(1)-O(5) | 1.95 | | 1.93 | |
| 2×V(1)-O(1) | 1.88 | 0.04 | 1.90 | 0.00 |
| 2×V(1)-O(3) | 1.92 | | 1.90 | |
| 3×$M$(2)-O(2) | 2.01 | 0.05 | 2.00 | 0.06 |
| 3×$M$(2)-O(1) | 2.06 | | 2.06 | |
| 3×$M$(3)-O(3) | 1.92 | 0.25 | 1.93 | 0.22 |
| 3×$M$(3)-O(2) | 2.17 | | 2.15 | |
| 3×Fe(4)-O(2) | 1.78 | | 1.77 | |
| Fe(4)-O(4) | 2.05 | 0.33 | 2.05 | 0.34 |
| Fe(4)-O(5) | 2.38 | | 2.39 | |

TABLE II. Wycokff positions and corresponding valance states of V for $AV_6O_{11}$ ($A$ = Na, K, Sr, Pb)[21,30] compounds, and also valance states of V/Fe for the PFVO obtained from Bond valence sum calculations in the present study.

| Site | Wycokff positions | Na/K$V_6O_{11}$ | Sr$V_6O_{11}$ | Pb$V_6O_{11}$ | PbFe$_{1.75}$V$_{4.25}$O$_{11}$ (BVS calculation) |
|---|---|---|---|---|---|
| $M$(1) | 6c | $V^{3+}$ | $V^{3+}$ | $V^{3+}$ | $V^{4+}$ |
| $M$(2) | 2a | $V^{4+}$ | $V^{4+}$ | $V^{4+}$ | $V^{3+}$/$Fe^{3+}$ |
| $M$(3) | 2a | $V^{4+}$ | $V^{4+}$ | $V^{4+}$ | $V^{3+}$/$Fe^{3+}$ |
| $M$(4) | 2b | $V^{4+}$ | $V^{3+}$ | $V^{3+}$ | $Fe^{3+}$ |

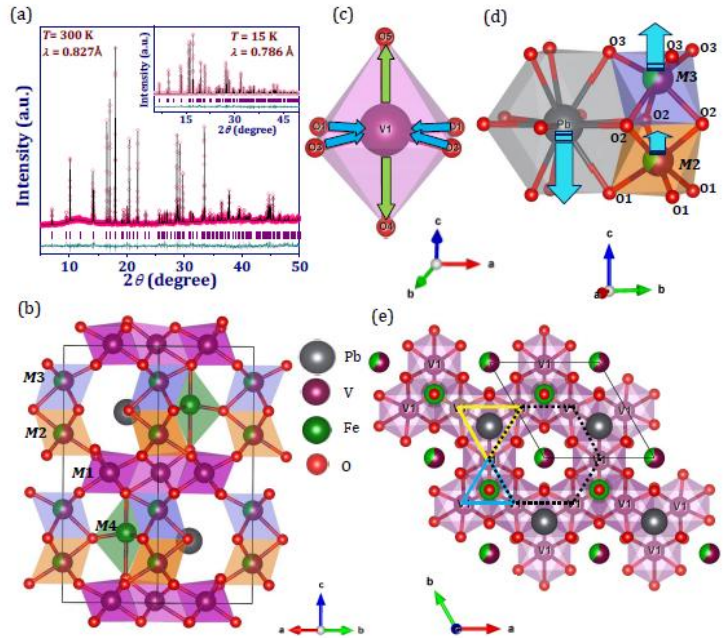

**Figure 1:** (a) Rietveld refined 300K XRD pattern of PFVO, inset: 15 K XRD data with fitting. Pink open circles represent the experimental data and solid black line represents the calculated pattern; The dark cyan line represents the difference between the observed and calculated pattern, while purple vertical lines signify the positions of the Bragg peaks; (b) Refined crystal structure. (c) a single $V(1)O_6$ octahedral unit with elongated V(1)-O bonds along z-axis; (d) $M(2)M(3)O_9$ Face-shared dimer units connected with $PbO_{12}$ cuboctahedron; Movements of the cation centers in the respective polyhedra are further shown by cyan arrows; (e) Kagome network formed by edge shared V(1) octahedra in the a-b plane; two distinct corner-shared equilateral V(1) triangular motifs, designated by yellow triangle with longer V(1)-V(1) distances and sky blue triangle with shorter V(1)-V(1) distances.

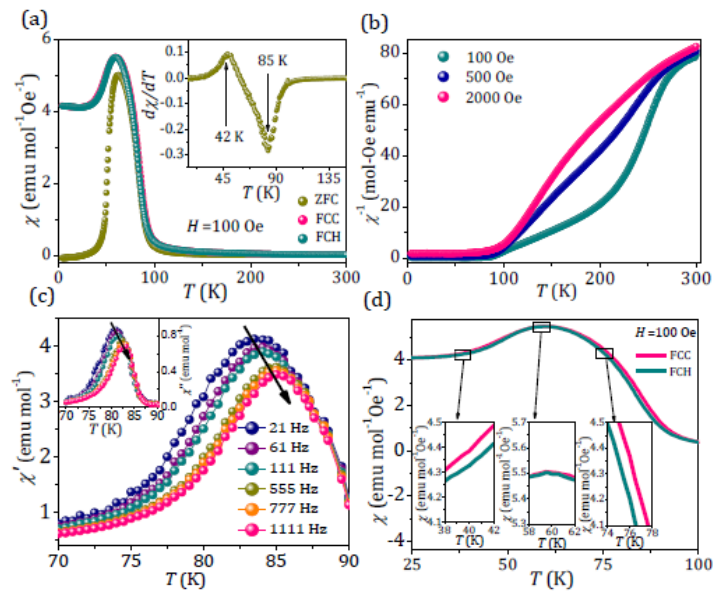

**Figure 2:** (a) Zero-field-cooled (ZFC), field-cool-cooling (FCC) and field-cool-heating (FCH) dc magnetic susceptibility variations as a function of temperature $\chi(T)$ under 100 Oe applied magnetic field; Inset: respective 1st order temperature derivative of the FCH susceptibility variation; (b) Temperature dependent inverse magnetic susceptibility variations for three different magnetic fields; (c) The temperature dependence of the in-phase ac susceptibility component $\chi'(T)$ at the different applied frequencies; Inset: the respective out-of-phase susceptibility $\chi''(T)$ variations. (d) Zoom-in view of the temperature dependent 100 Oe FCC and FCH susceptibility variations, indicating thermal hysteresis between the heating and cooling data.

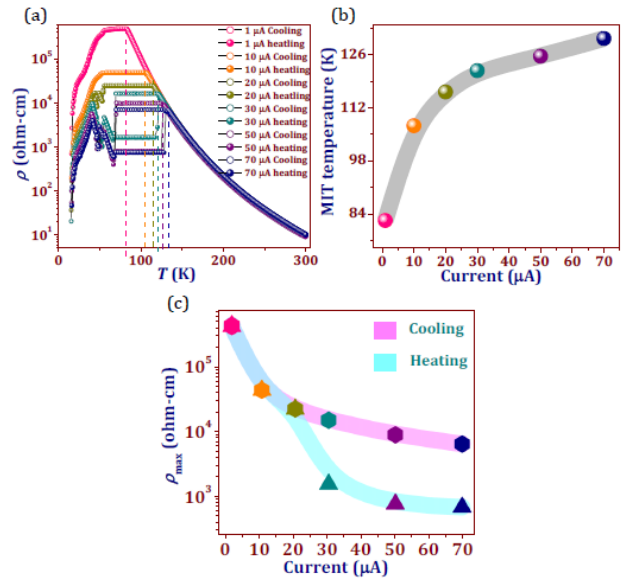

**Figure3:** (a) Temperature dependent electrical resistivity at the several applied electric currents; (b) Evolution of Metal-insulator transition (MIT) temperature with applied current; (c) Maximum value of resistivity at different applied currents during heating and cooling cycle.

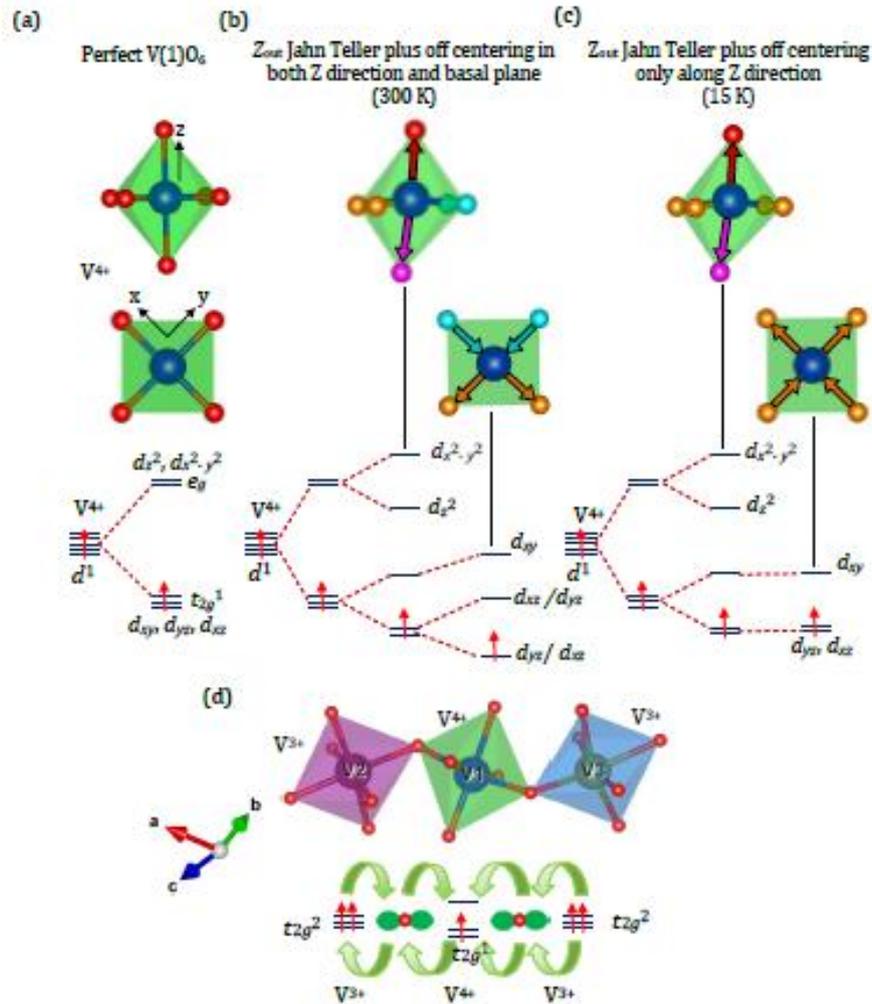

**Figure4:** (a) Perfect V(1)O$_6$ octahedra and corresponding energy level diagram. (b) $Z_{out}$ Jhan-teller distortion along Z axis and off centering movement of V(1) along Z-axis and within basal plane at 300 K, and the derived energy levels. (c) $Z_{out}$ Jhan-teller distortion along Z axis and off centering movement of V(1) only along Z-axis at 15 K with the corresponding energy level diagram. (d) Double exchange mechanism between V$^{3+}$ (V(2)/V(3) site) and V$^{4+}$ (V(1) site) via oxygen.